# NEW CONFIGURATIONS OF THE INTERFACE BETWEEN INNOVATION AND URBAN SPATIAL AGGLOMERATION: THE LOCALIZED INDUSTRIAL SYSTEMS (LIS) OF THE CLOTHING IN FORTALEZA/BRAZIL

Edilson Alves PEREIRA JÚNIOR
University of the State of Ceará (UECE)/Brazil
Doctor and Master Program in Geographic Studies (PROPGEO)

**Abstract**
The paper seeks to interpret the interface between innovation, industry and urban business agglomeration, by trying to understand how local/regional entrepreneurs and social agents use the forces of agglomeration to organize productive activities with a view to achieving greater competitiveness and strategic advantages. There are many territorial manifestations that materialize from this new reality and the text seeks to propose a reading of its characteristics based on what we call "productive spatial configurations", which represent the specific functioning of a certain innovative production process and its territorial impact. To illustrate this approach, we take as an example a case study that illustrates how productive spatial configurations manifest themselves through the revitalization of an industrial economy that incorporates innovative efforts, whether technological, process or organizational. This is the localized industrial system (LIS) of clothing and apparel in Fortaleza, Ceará state, Brazil, which reveals an industrial experience of resistance with significant organizational innovation in the production and distribution processes of clothing. The main objective of the proposal is to organize theoretical and empirical tools that will allow us to read the combination of economic, social and political variables in spaces where collaborative networks of companies, service centers and public institutions flourish, in the context of various industrial production processes. This could point to the progress we need to overcome the many false controversies on the subject.

**Keywords**: Innovation; Localized industrial system; Fortaleza, Brazil.

## 1 – INTRODUCTION

The text aims to interpret the interface between innovation, industry and space by trying to understand how diverse economic, social and political agents use the forces of agglomeration and the circulation of ideas and technology to organize productive activities with a view to obtaining greater competitiveness and strategic advantages.

The last few decades have seen an undeniable transformation of the experiences between innovation and business activities, with significant effects on industrial systems, which have reacted by making a transition in production engineering practices based on a greater territorial division of the stages of manufacturing goods. Thus, growing innovation, incorporated into industrial production arrangements, was responsible for creating different articulations between spatial, productive and service relations, capable of revealing a new wave of local and regional reindustrialization.

These are dynamics that, in combination with each other and the space, generate a set of economies or diseconomies associated with the agglomeration itself, marked by an atmosphere that produces a specific industrial landscape to innovation, defining a concentration and dispersion of production processes.

The text, far from confusing innovation with technological novelty, defends the idea that innovative processes is the result of dynamic interactions of collective learning. It also argues that innovative actions depend on a circular articulation which have systems of interactions based on economic regulations, political regulations, institutional initiatives and educational practices which, by interacting, contribute to the conception, manufacture and commercialization of something new, only effected in a social way.

There are many spatial manifestations that materialize from this new reality, and the article seeks to propose a reading of its characteristics based on what we call productive spatial configurations, which represent the specific functioning of a certain innovative production process and its territorial repercussion. The reading of these configurations is anchored in parameters that can be defined from multiple elements, including the origin and size of the industrial capital installed in a given region, the technological content used, the organizational productive renovations, the range of the spatial production circuits of the productive branches, the role of the State in making the entrepreneurships effective, the work relations present in the factories/workshops/offices and the specific spatial forms of the agglomerations of establishments.

To illustrate this approach, we take as an example a case study in which the relationship between innovation and industry is very relevant. This is the localized clothing and apparel industrial system in the city of Fortaleza, state of Ceará, Brazil, which reveals a resistent industrial experience with significant organizational innovation in clothing production and distribution processes.

In addition to this introduction, the section two of the chapter discusses the context of innovation in recent decades and assess the impact that this perspective has on an atmosphere or industrial ecosystem of innovation, in combination with the interaction of the agents, the fluidity of investments and the renewal of technological and productive aspects. In section three, it deals with the aforementioned empirical manifestations based on the notion of productive spatial configuration and the account of the localized clothing and apparel industrial system in Fortaleza, when the debate is lead to its final remarks.

## 2 – ECOSYSTEM AND THE MIDDLEGROUND INOVATION

Numerous studies have dealt with how innovation provides the appropriate context to explain significant transformations in places and regions (Shearmur; Carrincazeaux; Doloreux, 2016; Caravaca; Romero; Méndez, 2002; Castells; Hall, 2001; Storper, 1995). Among other grounds, it is assumed the premise that a company cannot develop a technical or organizational novelty solely in an abstract relationship with the market, without a whole atmosphere or ecosystem being conceived from the firm's relationship in an environment conditioned by several agents and their productive practices. Aydalot's (1986) classic expression is that if there is innovation, it depends first and foremost on the environment in which the firm evolves.

This is to mention the promising role of the articulation between a number of agents in environments where collaborative networks of companies, service centers, science and technology parks, public institutions and social organizations densify, which are responsible for fostering an atmosphere of creativity and discoveries of this initiative through collective learning processes.

However, differently from the traditional approach on the formation of innovation ecosystems (Moore, 1993), we understand that the networks of collaborations that lead to the consolidation or resilience of certain regions, even when they revitalize the productive economy and incorporate innovative effort, do not necessarily feature systemic interactions carried out by agents from a formal and consolidated economy. For the traditional approach, all innovative action will lead to a virtuous cycle of development, change of qualitative regional structural, technological improvement and higher pay/qualification of labor.

Even though such experiences usually produce an innovative capacity that boosts a considerable number of production and service companies, generating value added and with the capacity to drive a regional economy, we do not Always incorporate new qualities, especially when the combination adopts a competitiveness agenda based on cost reduction and predatory competition (Hassink, 1997). When this happens, innovative processes are responsible for renewing the system, but they also interrupt a virtuous cycle of technical and informational improvement, with detrimental effects for the majority of companies and people.

Among the references to address the issue, the works of Cohendet (2018) and Cohendet, Grandadam and Simon (2010) are considered, for whom, in addition to the interactions between agents of the formal and structured economy, the essential for the consolidation of an innovation ecosystem resides in the several informal active units that relate to structured groups, such as associations or unregulated communities, individuals or groups of informal service suppliers, small subcontracted producers, etc.

For the authors, what guarantees the renewal and resistence of the system is the middleground (Cohendet, 2018; Cohendet; Grandadam; Simon, 2010), which ensures the necessary combination of formal elements (firms, various organizations, research laboratories, etc.) and informal elements (various communities, individual actions, independent groups, etc.). According to Cohendet (2018, p. 98), the middleground is "conceived as a set of mechanisms and intermediate combinations that allow to connect formal organizations with individual actions or informal collectives, and give the ecosystem its generative power, resilience and attractiveness". The Figure 1, adapted from Cohendet (2018), illustrates the agents and their set of combinations.

The middleground consists of four different elements that ensure the combination of the dynamics of informal actions and agents confronted or adapted to the formal structures of the markets. They are: 1) places; 2) cognitive interaction environments and platforms; 3) events; and 4) projects (Cohendet, 2018).

We can describe them briefly (Cohendet, 2018):

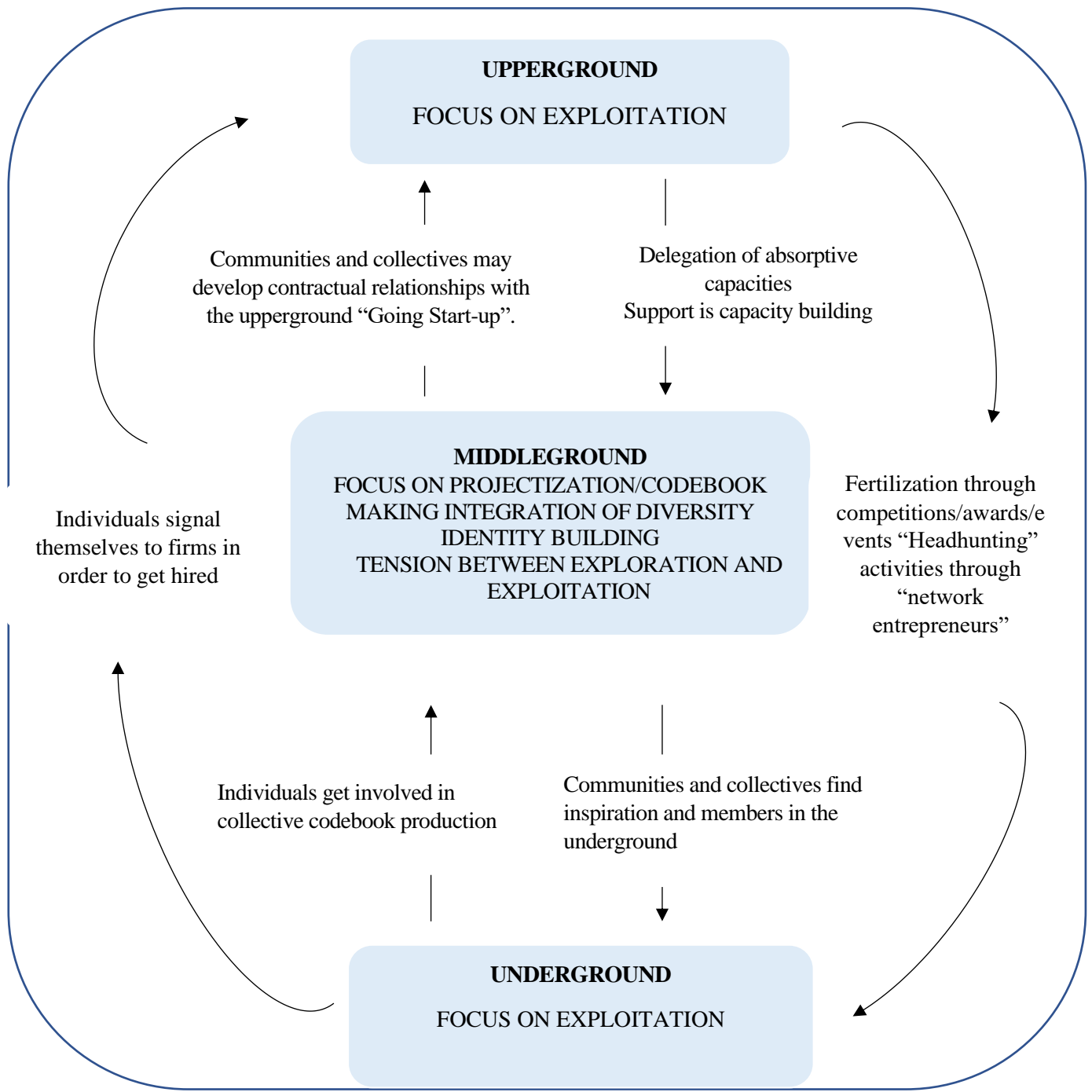

**Figure 1. Upper, middle and lower territorial devices (Upperground, middleground and underground)**

1) the meeting and relationship places established by agents of innovation are a fundamental geographical dimension that receives and stimulates the exchange of experiences, knowledge and business, on the border between formal and informal relationships, in an explicit or tacit way. They provide the conditions for heterogeneous groups to meet (sales, conventions, teaching and research centers, etc) without which the atmosphere of interaction would be compromised;

2) cognitive interaction environments and platforms allow different communities to interact and structure ideas, especially through networking and the exchange of information and technology. Different experiences are tested and approved/disapproved as a result of these articulations, which can assume a trans-scalar relationship by linking the transfer of information locally, nationally and globally, transcending the limits imposed by proximity;

3) the events allow agents and formal organizations to stimulate their cooperation and competitiveness through exhibitions, but also to detect new ideas, talents, organizational systems and experiences with individuals or groups from different domains. Although many street markets, releases and exhibitions are a formal innovation and publicity strategy, the results of these meetings generally reach the lower territorial level (underground) of the innovation ecosystem, involving the informal elements, even if by predatory exploitation;

4) finally, projects involve, on temporary occasions, the engagement of different members from comunitites in a context of a common vision and experience. In general, the role of public institutions and the State is fundamental in the conception and prospecting of these projects, and they often mediate agents of different scales and different links, stimulating integration ties that can be ephemeral or long-lasting.

The middleground, therefore, ensures its important capacity of combining elements and agents, playing a role of: (a) a permanent mechanism of innovation, association and formalization of heterogeneous knowledge basis; (b) means of detecting and seizing ideas, instruments or organizational experiences that have arisen in the underground of the innovation ecosystem; (c) platforms that create, recreate and energize productive, economic and social relations and (d) devices of interaction and renewal between the underground and the upperground, allowing the informal and the formal not only dialogue, but renew themselves, one in interaction with the other (Cohendet, 2018).

However, in the relationship between the devices that allow a more comprehensive understanding of the environment and the industrial atmosphere of innovation with the geographical reality produced by concrete agents, we believe that the spatial context of the several agents and their productive practices go through economic, political, cultural and social expressions manifested in the territory. That is to say, the abstract concept of an innovation environment or ecosystem, when linked to economic and industrial activity, is insufficient if it does not take into account concrete contexts, national, regional or local features, geographical implications of the differentiations, frequencies, interactions and limits outlined in space.

This is why we propose an approach closer to objective spatial experiences and that can be read from defined geographical categories, which will be discussed in the next section.

# 3 – SPATIAL EXPRESSIONS OF INNOVATION IN INDUSTRY: AN APPROACH OF SPATIAL CONFIGURATION OF PRODUCTION

We believe that one way of interpreting the relationship between industrial geography and innovation can be conducted by what we call productive spatial configurations, which represent the specific functioning of a certain innovative production process and its territorial repercussion. The materialization of these configurations is based on parameters that can be defined from multiple characteristics, including the origin and size of the industrial capital installed in a given region, the technological content used, the organizational productive renovations, the range of the spatial circuits of production of the productive branches, the role of the state in consolidating entrepreneurships, the work relationships present in the factories/workshops/offices and the specific spatial forms of the agglomerations of establishments (Pereira Júnior, 2015).

All these characteristics, combining each other and the place, tend to generate a set of economies or diseconomies associated with the agglomeration itself, marked by an atmosphere that produces a specific industrial landscape of innovation, defining a concentration and a dispersion in terms of the spatial dynamics of production processes.

These productive spatial configurations are also characterized by the relationship established between proximity and geographical discontinuity, which demarcates the divisions between political regulatory systems, intensity of technologies, competitive advantages, governance proposals, actions of economic agents, etc., being the truly responsible for delimiting the different spatial areas consolidated by industrial activities. In the same way, they articulate different scales, showing the interdependence of the agents involved in the process and the flows they trigger.

As examples of productive spatial configurations, we include technology parks, industrial condominiums, localized industrial systems, industrial complexes, industrial zones, industrial axes, and so on, where institutional relations (intercompany, state-owned, etc.), a number agents and infrastructure are materialized as principles of geographical orientation to promote efficiency and the constitution of a productive coordination framework.

Industrial activity has traditionally been one of the main conditioning factors for the expansion and organization of space, as well as establishing a two-way relationship with it, either by taking advantage of the numerous benefits of the agglomeration of people and services, or by stimulating the assembly of infrastructure and the circulation of goods and information. Currently, increasingly subjected to a diffuse network of territorial relations that integrates people, information, goods and investments, industry is incorporating changes directly linked to both the scientific content of its production engineering and the spatial scope of its consumer market. These changes engender quantitative and qualitative transformations, affecting the distribution of establishments in places with a tradition of production or in peripheral areas.

We see innovation as an important dimension of all these transformations, insofar as it incorporates new creation practices that can favor industrial arrangements through the incorporation of technologies, greater economic diversification, improvement of productivity, job quality and higher payment for workers. However, when it comes to industrial spaces, an innovative process is not always marked by the incorporation of new qualities. When industry absorbs the innovative influence of productive and technological interactions and adopts a competitiveness agenda based on cost reduction and predatory competition, it interrupts a virtuous cycle of technical and informational improvement. In this case, the structural change is not qualitative, although the innovative processes are clearly present as a strategy of the different agents who interact to conceive a new element for the system.

To illustrate such perspectives, we have taken as an example a case study of the localized clothing and apparel industrial system in Fortaleza, state of Ceará, Brazil, which reveals an industrial experience of resistance with significant organizational innovation in clothing production and distribution processes.

We must not forget that this empirical section, in order to be properly contextualized, needs to consider two important aspects, without which its concreteness cannot be understood. Firstly, it must overcome the fetishizations of innovation that have already been announced, especially the one that gives the impression that everything that is innovative is of superior quality and, at the same time, represents invention and the spread of high technology.

Secondly, it is a spatial configuration that represents an empirical section of the relationship between industry and innovation in a Latin American country, i.e, Brazil, which in the last five years has been going through one of the most troubled periods in its political history, which has reverberated in the economy and, of course, in industry. In order to mention this important fact without prolonging the discussion, we should remember that among the new phenomena materialized in Brazilian industry is the fact that the country has assumed four movements in its industrial project, namely: a) the defense of competitiveness through reduction in costs measures such as relocating factories, increasing flexibility in labor relations and reducing wages and social security charges; b) a potencialization of entrepreneurial profitability via tax exemption between different regions and places in the country, with encouragement for "tax wars"; c) a complementarity to international productive capital, acting domestically in activities subordinated to maquila; and d) a tendency to concentrate on productive genres that offer competitive advantages associated with the extraction and processing of natural resources.

For the purpose of this text, we are interested in the fact that Brazilian industry has abandoned its ability to reproduce itself through technological refinement and innovative autonomy, assuming a subordinate position in relation to global value chains.

## 4 - THE LOCALIZED INDUSTRIAL SYSTEM OF CLOTHING AND APPAREL IN FORTALEZA

Localized industrial systems (LIS) represent clusters of productive establishments with specific geographic patterns, with a strong aspect of local rooting and the dynamization of network relationships (Courlet, 1993; Pecqueur,

1993; Piore & Sabel, 1986). They are not configured in the same way everywhere, since each manifestation has its own features and socio-economic conditions related to the predominant regional experiences of reproduction.

They can also provide strong intermediary responses between the local and the global, as they are often articulated through diffuse industrialization, which reinforces their reticular character and at the same time does not exclude traditional forms of productive organization, such as home-based work, artisanal work and subcontracting networks, which are denser (Zimmerman, 2000). Thus, LIS is characterized as a productive spatial configuration that frequently combines the flexibility of small networks and agents operating on different scales (Courlet, 1993; Pecqueur, 1993).

The role of the agents should be highlighted, especially due to their socio-economic ties, with emphasis on public agencies, associations of entrepreneurs and workers, training/educational institutions and private companies. These relationships may reduce uncertainties in the process of forming and consolidating agglomerations, especially in the area of innovation and its relationship with the organizational functioning of the industry in the territory. When this relation does not happen, LIS does not have the capacity to be competitive and is vulnerable to exogenous forces, which disqualifies the main existing regional possibilities, often implying a predatory relationship between local agents (Hassink,1997).

In line with what Cohendet (2018) suggests, these interactions between agents should be highlighted. They take place through political, cultural and socio-economic relations and are responsible for the internal and external articulations of the system. However, in addition to the interactions between agents of the formal economy, the strength of the system and its resistance also lies in the numerous informal active units that relate to structured groups, such as individuals or groups of informal service providers, small subcontracted producers, etc.

Through internal and external articulations, it is necessary to understand LISs based on a synergy of behaviours and an innovative organizational capacity to articulate companies and their competitive strategies, ensuring the necessary combination of formal elements (companies, patronal syndicates research laboratories, etc.) and informal elements (workshops, informal home work, etc.), resulting in an impact on competitiveness, productivity and, consequently, the system's resistence. Similarly, local policies must be specific to each industrial system, since even though they have common characteristics, they have particularities linked to the productive and the local sector.

Fortaleza's clothing and apparel LIS is used in this text as an empirical expression of the relationship between innovation, industry and spatial arrangements[1]. It represents the worsening competitiveness of the clothing and apparel industry in Brazil, which since the 1990s has felt the effects of a territorial and productive restructuring incorporated through cost reduction measures, such as relocating factories, increasing flexibility in labor relations and reducing salaries.

---

[1] The clothing and apparel SIL in Fortaleza specializes in beachwear and underwear and began its production as a result of the expansion of the textile industry, which took place between the end of the 19th century and the 1980s. The clothing industry was strengthened in the 1990s and consolidated in the following decades, when it began to serve the foreign market. Between 1990 and the current period, changes in production and organizational innovation have been tested and implemented.

It is an industrial production agglomeration located in the metropolitan area of Fortaleza, in the state of Ceará, in the Northeast region of Brazil. While the gross value of industrial production of clothing and accessories in Brazil in 2018 reached a total of US$11 billion, Ceará accounted for 6.21% of the country's production in the period, which corresponded to 48.1% of everything produced in the Northeast region (IBGE, 2022).

Given the production in the state of Ceará, the clothing and apparel LIS in Fortaleza metropolitan area is quite relevant. According to 2021 data from the Brazilian Ministry of Labor and Employment (MTE/Brazil), through the Annual Social Information Report (RAIS, 2023), the agglomeration, which includes the municipalities of Fortaleza, Maranguape, Maracanaú, Caucaia, Horizonte, Pacajus, Cascavel, Aquiraz, Chorozinho and Eusébio, had 33,155 formal jobs and 1,881 production establishments in this industry. This was equivalent to 83% of Ceará's total in both indicators, which in the same year concentrated 40,035 jobs and 2,262 formal production establishments (see Table 1).

As this is a metropolitan location, the companies established there prefer to cover the costs of the diseconomies of scale of the big city and its region of influence, avoiding moving their factories to more remote environments. Of course, proximity to an abundant and cheap workforce and a thriving consumer market has a significant influence on the profitability goals of these companies, resulting in specific spatial forms of an industry that is distributed in residential and commercial areas, realigning production stages according to a new innovative organizational engineering.

This new engineering makes production on the factory floor more flexible and redistributes it throughout the city and metropolitan area, making circuits more dynamic through the relocation of tasks, functional disjunction, externalization of production and the use of spatial differences, with the aim of reducing costs, increasing productivity and reducing obstacles to profitability.

| Municipalities | Nº of formal productive establishments | Nº of formal works originated |
|---|---|---|
| Fortaleza* | 1.566 | 22.432 |
| Caucaia* | 82 | 1.255 |
| Maracanaú* | 77 | 4.042 |
| Maranguape* | 57 | 2.058 |
| Juazeiro do Norte | 47 | 360 |
| Sobral | 39 | 511 |
| Frecheirinha | 37 | 2.425 |
| Aracoiaba | 34 | 575 |
| Aquiraz* | 23 | 136 |
| Cascavel* | 21 | 469 |
| Pacajus* | 20 | 847 |
| Pacatuba* | 18 | 1.104 |
| Horizonte* | 17 | 668 |
| Total for the State of Ceará | 2.262 | 40.035 |

**Table 1 - Number of production establishments and formal jobs in the clothing and apparel industry in the state of Ceará/2021**

* Municipalities belonging to the clothing and apparel SIL in the Fortaleza metropolitan area.
Source: Annual Social Information Report (RAIS, 2023). Brazilian Ministry of Labor and Employment (MTE/Brazil).

This strategy could not be implemented without the urban space being used as a mechanism to facilitate these relations of production and consumption, precisely because of the articulations and asymmetries that characterize it. The segment is predominantly made up of micro, small and medium-sized clothing production companies, which are organized in the form of garment factories and "facções"[2], i.e. sub-units contracted to manufacture parts or the complete garment (see Table 2).

Although they are scattered throughout all the metropolitan municipalities, they are predominantly located in neighborhoods with lower family incomes, as Figure 2 shows. They are workshops located in the their owners homes, occupying idle rooms, garages, sheds, among others, apparently invisible to those who walk the streets of the neighborhood.

---

[2] The "facções" are workshops that provide services to other factories at various stages of production. They can either specialize in a particular stage or produce the entire garment.

| Municipalities | Nº of formal productive establishments | | | | Nº of formal works | | | |
|---|---|---|---|---|---|---|---|---|
| | Micro | Small | Medium | Big | Micro | Small | Medium | Big |
| Fortaleza* | 1.290 | 191 | 19 | 2 | 6.297 | 7.944 | 3.310 | 4.831 |
| Caucaia* | 60 | 10 | 2 | 0 | 260 | 412 | 467 | 0 |
| Maracanaú* | 55 | 14 | 2 | 2 | 192 | 701 | 657 | 2.459 |
| Maranguape* | 39 | 16 | 2 | 1 | 233 | 754 | 223 | 833 |
| Aquiraz* | 19 | 3 | 0 | 0 | 62 | 68 | 0 | 0 |
| Cascavel* | 12 | 5 | 1 | 0 | 45 | 249 | 118 | 0 |
| Pacajus* | 14 | 4 | 0 | 1 | 57 | 179 | 0 | 608 |
| Pacatuba* | 10 | 5 | 0 | 1 | 41 | 223 | 0 | 824 |
| Horizonte* | 11 | 2 | 1 | 1 | 56 | 54 | 165 | 0 |
| Total for the State of Ceará | 1.774 | 312 | 39 | 8 | 8.525 | 12.976 | 6.910 | 10.164 |

**Table 2 - Production establishments and formal jobs by company size in the clothing and apparel industry in the Fortaleza metropolitan area/2021**

* Municipalities belonging to the clothing and apparel SIL in the Fortaleza metropolitan área.
Source: Annual Social Information Report (RAIS, 2023). Brazilian Ministry of Labor and Employment (MTE/Brazil).

They hire an unskilled workforce, living in the neighbourhood itself, mostly unformalized and responsible for producing a large volume of uniforms, underwear, beachwear, jeans and other garments made from light fabrics. Although the total number of workers employed per establishment is usually small, the high number of workshops makes possible to employ a significant number of workers, probably three or four times higher than the formal indicators in Table 1, as we can see from the interviews with well-informed agents and the leaders of the associations in this branch of production.

**Figure 02 - Location of formal jobs in the clothing and apparel industry in the Fortaleza LIS/ 2021**

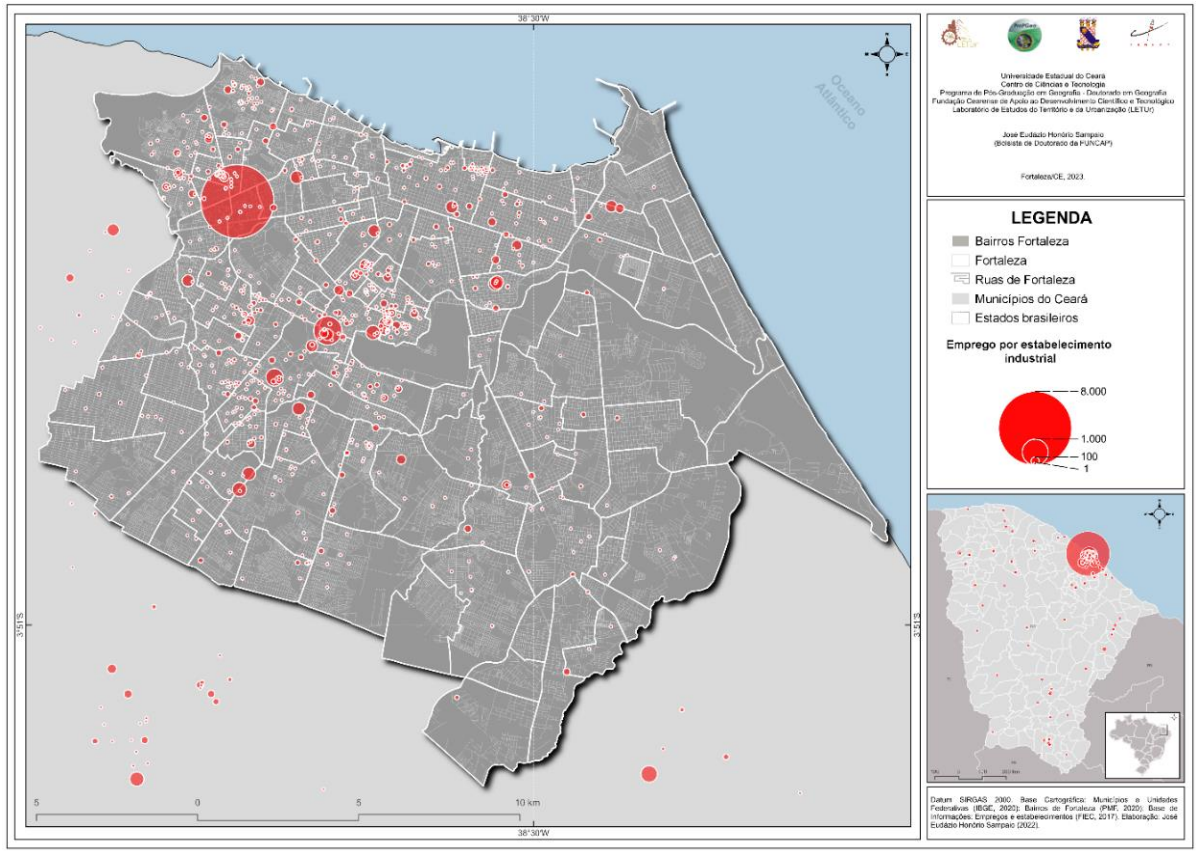

Source: Annual Social Information Report (RAIS, 2023). Brazilian Ministry of Labor and Employment (MTE/Brazil). Federation of Industries of the State of Ceará (FIEC)/2021.

This increases the spatial division of production and labor within the LIS, while at the same time leading to a precarization of the services provided, but it also highlights the role of interaction between the underground and upperground territorial devices of this productive spatial configuration, ensuring, as Cohendet (2018) informs us, the strategic capacity of the middleground territorial device in allowing the informal and formal sectors not only to dialogue, but also to renew themselves, one in interaction with the other (Cohendet, 2018).

The problem with the relationship between the aforementioned territorial devices is that the creation of specializations, the integration of diversity and the construction of productive innovations takes place under strong tension of exploitation between the agents. While some mediating devices between the higher and lower devices insist on the idea of cooperation, such as educational institutions and cooperatives, the greatest intermediary force of articulation in Fortaleza is a tense relationship between companies, syndicates, public authorities and precarious workers.

As the territorial devices are unable to engage in a dialog around a cooperative objective, spurious competition and exploitative relations predominate, with perverse effects for small producers and production line workers, especially those hired for informal jobs or who work from home.

Thus, the resilience and reproduction capacity of this LIS reveals factors that vary according to the strategic capacity of the middleground, especially when it is led by companies that act as anchors, i.e. they are always ready to leave if they no longer find advantages in staying in a particular place (Zimmerman, 2000).

These factors include a) the relationship between industrial production and urban space in the metropolis of Fortaleza, with focus in neighborhoods with a low-income population, on the periphery of cities, in houses and backyards, boosting the productive spatial circuit mainly through micro and small establishments, usually informal; b) the relationship with street markets located in the city and in the metropolitan area of Fortaleza, for marketing, such as local and regional fairs, but also the contribution to the tourism sector, which is quite expressive in the state of Ceará; c) intensive use of workforce, especially female labor (Femicro, 2017); d) low investment capital and low technology, with manual labor still being essential (Femicro, 2017); e) the high level of informality in the sector, due to the low barriers to entry and the availability of labor for middle activities, such as cutting threads and garments; f) the relevance of the economic activity for the region and the state, which guarantees its profitability margins even with international competition, especially from Chinese products (ABIT, 2018).

This articulation highlights the already discussed firm-territory dialectic by Zimmerman (2000), which makes companies effective agents of territorial dynamics. This is materialized through network of companies, attached to communication techniques based on fast relationships. Vertical structures disintegrate and are articulated in a network of establishments specializing in certain production stages, which means aligning micro and small establishments or residential work done by women and children.

These new organizational patterns are only succeeded through the use of technology capable of employing highly advanced machines, which use the CAD/CAM[3] system to cut the garments and which make use of hydraulic rocker arms, even though most of the small producers and home workers use machines scrapped or cut and sew the footwear manually.

This clothing and apparel LIS then has a specific spatial form that materializes in Fortaleza, marked by the expansion of investments, forms and industrial flows to peripheral districts of the municipalities in the metropolitan area, based on an example in which the organizational innovation of production processes manages to overcome the barriers imposed by the precarization of transport infrastructure, the low qualification of the workforce and the dispersion of production establishments.

---

[3] CAD (computer aided design) and CAM (computer aided manufacturing) are programs that allows the creation and the programming of projects using a computer. The programmer designs instructions and product cuts and specifications that are controlled based on a predetermined design.

Finally, as a result of this spurious competition, which reduces the price of labor, makes production conditions more precarious, does not incorporate innovative actions for the benefit of the majority of producers and expands the informal relations of the productive arrangement, some companies are inserted into national and global value networks subordinately. As an example, we can consider that medium-sized companies, and even some large ones, are unable to compete both in and outside the country with their own clothing brands, preferring to acquire production licenses from large retail brands.

By assuming this position, which neglects the autonomy of the production and branding of the goods traded, producers are subordinated to the instabilities of the market and the decisions of hollow corporations (Michalet, 2009), being coordinated by external agents who are often indifferent to the localized system. This does not benefit the innovation environment, especially as it follows a path of insertion based exclusively on the lowest cost of the product, through spurious organizational reengineering that reproduces strategies of exploitation between agents.

## 5 - CONCLUSION

The productive spatial configurations approach helps to understand the combination of innovation, industrial production and geographical space, since it starts from a set of common variables in terms of economic, social, political, cultural and territorial dimensions to adress specific cases and spatial forms, i.e. empirical experiences found in the concrete world. It strengthens the idea that innovation is capable of explaining the dynamics of certain regions or places based on the correlation between agents and their specific functioning on a material level.

This is why it is often the case that even when dealing with an effective combination of technological breakthroughs, scientific synergy, organizational recomposition and economic productivity, a certain productive spatial configuration will only point to prodigious strategies for most members of a region or place if the political regulatory systems, the use of technology, the actions of socio-economic agents and governance proposals do not establish spurious relations of exploitation in the innovation environment.

Unveiling the virtues or unsuccessful experiences of these relationships between innovative actions and their procedural and organizational use involves what has already been pointed out in the text as circumstantial in understanding the meanings of innovation, that is, it results from dynamic interactions of collective learning in a context of systems of interactions only carried out in a social way.

This makes it easier to determine whether or not an innovation environment is promising for the articulation of different agents with cooperative strategies. It is also possible to find out if collaborative networks between companies, institutions, public authorities and various representations really do foster an ecosystem of creativity and discovery through collective learning processes.

The experience of the case study, the Localized Industrial System for clothing and apparel in Fortaleza, in the Northeast of Brazil, illustrates how productive spatial configurations manifest themselves through the revitalization of an industrial economy that incorporates innovative efforts, whether technological, process or organizational. However, these innovations may present different content, engagement and systemic interactions, not necessarily leading to a virtuous cycle of development, qualitative regional structural change, technological improvement and higher pay/qualification of the workforce.

Comprehensively organizing the theoretical and methodological tools that make it possible to read the combination of economic, social and political variables in spaces where collaborative networks of companies, service centers and public institutions grow, in the context of diverse industrial production processes, may help uncover concrete challenges regarding the patterns of creativity that will lead to this virtuous cycle of development. This could point to the breakthrough we need to overcome the many false controversies on the subject.

## 6 - REFERÊNCIAS